# Survey study of the QoS Management in Distributed Interactive Simulation Through Dead Reckoning Algorithms

*Akram HAKIRI* [1, 2],
*Pascal BERTHOU* [1, 2],
*Thierry GAYRAUD* [1,2]

[1] CNRS ; LAAS, 7, avenue du Colonel Roche, 31077 Toulouse, France
[2] Université Toulouse; UPS, INSA, INP, ISAE; LAAS; F-31077 Toulouse, France
Email: {Hakiri, Berthou, Gayraud}@laas.fr

**ABSTRACT**: *Dead Reckoning mechanism allows reducing the network utilization considerably when used in Distributed Interactive Simulation Applications. However, this technique often ignores available contextual information that may be influential to the state of an entity, sacrificing remote predictive accuracy in favor of low computational complexity. The remainder of this paper focuses on the analysis of the Dead Reckoning Algorithms. Some contributions are expected and overviews of the major bandwidth reduction techniques currently investigated are discussed. A novel extension of Dead Reckoning based on ANFIS systems is suggested to increase the network availability and fulfilling the required QoS in such applications. The model shows it primary benefits regarding the other research contributions, especially in the decision making of the behavior of simulated entities.*

## 1. Introduction

Distributed Interactive Simulation (DIS) applications are sharing multiple entities in distributed in large scale networks. Updating these entities states generates a high quantity of information to be sent over physical link, which saturate the bandwidth of the underlying network.
In order to reduce the bottleneck and help to better manage the available network resources, Dead Reckoning (DR) algorithms were proposed as message filtering techniques. It is a process consisting in the estimation one's current position of an entity based upon a previously determined position and advance this position based upon known or estimated speed over elapsed time.

This paper suggests a mathematical formulation of the QoS requirements in DIS applications to be used in Dead Reckoning Algorithm, and an Adaptive Nero-Fuzzy Inference System (ANFIS) based model is. ANFIS Reckoning succeeds in reducing the spatial and temporal errors associated with remote entity model. It seeks to replace the use of instantaneous motion information with a predictive scheme in order to improve the spatial and the temporal coherence of remote entity. Consequently, a diminution of network traffic can be achieved through the decrease of frequently data transmitted due to the error threshold violation.

This paper is organized as follow: after a brief introduction in Section 1, an overview of the Dead Reckoning is given in Section 2. Section 3 presents some related alternative of Dead Reckoning mechanism. The reminder of section 4 is to discuss the QoS requirements of DIS applications. In Section 5 we introduce the mathematical formulation of the DR problem. Section 6 introduces the ANFIS Model for the DR algorithm. Conclusion is given in section 7.

## 2. Dead Reckoning Algorithm

For entities to interact meaningfully in distributed environment, coherence between entities' must be maintained. In order to reduce the number of updating message sent, Dead Reckoning technique was proposed. This technique dates back to the navigational techniques used to estimate ship's current position based on start position, travelling velocity and elapsed time.
DR is used to reduce the bandwidth consumption by sending update message less frequently and estimating the state information between updates.
Each remote site maintains in addition to the real representation of the entity, a high fidelity model to estimate the remote entity states locally. The anticipated entity states are computed from the last states based on the extrapolation of the position, velocity, rotation and acceleration of the entity.
When the gap between the extrapolated states and the real states exceeds a defined threshold ($TH_{pos}$ for the position and $TH_{or}$ for the orientation), the simulator transmits messages more frequently to anticipate the entities states motion. DIS standard [1] specifies a duration which is necessary to refresh the entities states with duration equal to HEART BEAT TIMER fixed to 5 seconds.

On all other the remote sites interested on this entity, the reception of new packets implies the updating of the entity states using first order or second order (equation (3)) extrapolation equations:

- Second order extrapolation: in addition to the given $P_i$ and $V_i$ from the first order extrapolation, the acceleration $A_i$ is added to the second order extrapolation. So, the quadratic extrapolation is given by :

$$P_{DR}(t) = P_i + V_i(t - t_i) + A(t - t_i)^2 \quad (1)$$

The flowing pseudo-code represents the second order extrapolation algorithm:

```
Inputs: S←State; t←time
Outputs: S' ← predicted state
D←t(s)
V←velocity (prediction (s)).
A←acceleration (prediction (s)).
Return: State (S') = state (S) +V*D+1/2*A*D²
```

Figure 1: second order predictive algorithm

The host that simulates the entity has to run a construction model (figure 2) in order to know how remote hosts perceive this entity.

```
DR-Msg (S,T,P, Q)
Inputs: S, Time Stamp (T), PredictedState (P),
previously sent Data (Q)
Outputs: New update Msg(Q), else NULL
CNST: Diff (TH-L)
If ((S-Prediction (Q,T)) < L) return NULL
Else
State (U) ←S;
Time (U) ←T;
Prediction (U)←P
Return (U)
```

Figure 2: reconstructed DR Message

## 3. Related work

There has been a tremendous amount of feedbacks from the users of the DIS applications regarding the DR tests. Authors in [2] present a dynamic filtering technique to send messages introducing a dynamic threshold denoted Update Lifetime (UL). UL is the delay between two consecutive updates and aims to make a new dynamic DR threshold. If UL is smaller, then an important threshold is produced to increase the performance of the communication. Otherwise, a small Threshold will be produced. Authors in [3] introduce also the same technique referred to Variable Threshold for Orientation (VTO). An auto-adaptive DR algorithm was presented in [4]. It fuses a dynamic multi-level DR threshold with the relevance filtering. It uses the distance between two simulated objects to calculate the threshold: a small distance involves a small threshold whereas a larger distance induces more important filtering of messages, which ensure the scalability of the simulation.

Two other adaptive algorithms are described in [5]. The first is centered on an adaptive adjustment of the threshold, whereas the second benefits from the previously determined values of TH to automatically select the extrapolation equation. However, it needs always data collected from the live simulation. Authors in [6] present a deterministic estimator of the objects parameters (position, velocity…). They use strict mathematic models to define deterministic events. Nevertheless, this model will generate a larger estimated error when an unpredictable behavior of the entities occurs. Kalman filters were also present in [7] to estimate the mobility of ad-hoc mobile nodes and then reduce the unwanted traffic. In fact, this approach provides 10% of the bandwidth optimization, and it was suggested in [8] for humanoid robot motion.

A fuzzy logic approach based on the fuzzy correlation degree of all the measured parameters of the entities (position, size, vision angle, velocity…) was proposed in [9], and it considers a multi-level threshold which applications should obey. A Nero Reckoning algorithm was suggested in [10, 13 and 14] as an intelligent approach to resolve the complexity of some sophisticated polynomial extrapolation. The remote host uses a Neuronal banc that includes the desired parameters (position, the velocity, the orientation…) to produce the new predicted proprieties of the simulated objects.

Through this panoramic vision, we presented several techniques and approaches discussed in the literature and concerning the network resources optimization and providing some answers to the QoS requirements in DIS exercises.

## 4. QoS requirements of DIS applications

Increasingly, DIS applications are not only constraints because their distributed and interactive nature, but also by cause of the requirements needed by the underlying network.

This cohesion is evident for coherent representation of entities and all occurring events. Two fundamental aspects have to be taken into account to ensure that cohesion: first, temporal coherence related to events, and then spatial coherence involves seeking the collection of entities states to a closed error.

The spatial coherence requires that at any time of the simulation, the gap between the entity state in the sender site $S_e$ and that in the receiver site $S_r$ does not exceed the threshold. For example, in figure 3, the gap of the position shift represented by $E_p$ should fulfill the following condition: $Th_{pos} \geq |E_{pos}|$ and the orientation error should satisfy the relation $Th_{or} \geq |Eor|$.

The temporal coherence involves that every remote site have to know the occurrence of all events occurring on all other sites within a bounded delay, for example an event occurring on the sender site must be seen on the receiver site after 100ms.

Furthermore, the QoS requirements are expressed using three terms:
- Reliability or the maximum allowable error packet lost, designated by $\tau$, which is intimately related to the maximum admissible error to ensure the spatial coherence of the simulation.
- Latency: maximum allowable delay across the network denoted DT, which is closely related to the temporal coherence.
- Jitter: the temporal deviation of this latency referred to us $\Delta DT$.

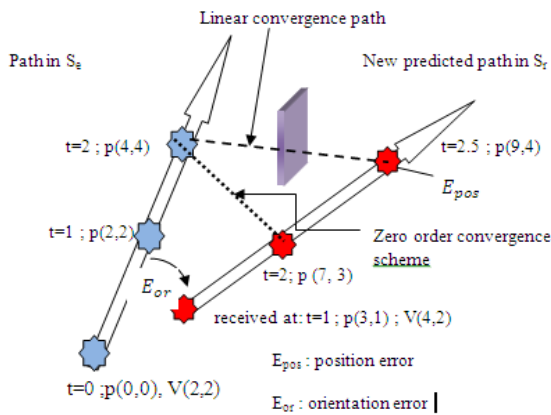

Figure 3: Dead Reckoning with prediction and convergence scheme

In [11], these parameters are specified in accordance with the coupling level between entities:

Tightly coupled interactions: this type of coupling may occurs when numerous entities are in narrow zone, and transmitted data packets require greater performances to ensure the coherence and the consistence of the simulation. in that case, latency is defined as $DT \leq 100$ ms and the error packet lost is given as $\tau \leq 2\%$.

Loosely coupled interactions: this type of coupling may occurs when simulated entities are fairly numerous and the distance which separates them is large enough to tolerate the transmission errors. Latency requirements between the output of data packet at the application level of a simulator and input of that data packet at the application level of any other simulator in that exercise is defined as $DT \leq 300ms$, and the error packet lost is given as $\tau \leq 5\%$.

In fact, the expression of the QoS regarding the coupling level has three major drawbacks:

- L1: the determination of the coupling level between distributed entities (which may evolve over time) seems to be expensive in terms of computation time.
- L2: the expression of the QoS based on the coupling degree (sender sites and receiver sites) sits uneasily with the multicast transport.
- L3: the DIS standard ignores the influence of the network latency on the position/orientation errors and neglect the spatial coherence constrains.

We will explain this limitation in the next paragraph. In order to simplify the presentation, we present only one dimension parameter of the simulated entity.

Let's consider figure 4, and assuming that Se and Sr are implicated in the same DIS exercise and exchange data packets including the position and the velocity of the entity, that we note here A. we will focus on the visible behavior of A assuming that the DR approach is applied to reduce the network traffic. Figure 4 illustrates the extrapolated error of the position of the entity A in both $S_e$ and $S_r$ sites, which we note respectively $E_s$ and $E_r$.

We chose the black circles to illustrate the packets sent from the sender site Se and the gray circles to indicate the reception of these packets at the receiver site Sr.

Dates Te0, Te1, Te2 and Te3 illustrate the date when packets are sent from site Se, and dates Tr0, Tr1, Tr2 and Tr3 indicate the date when packets are received on site Sr.

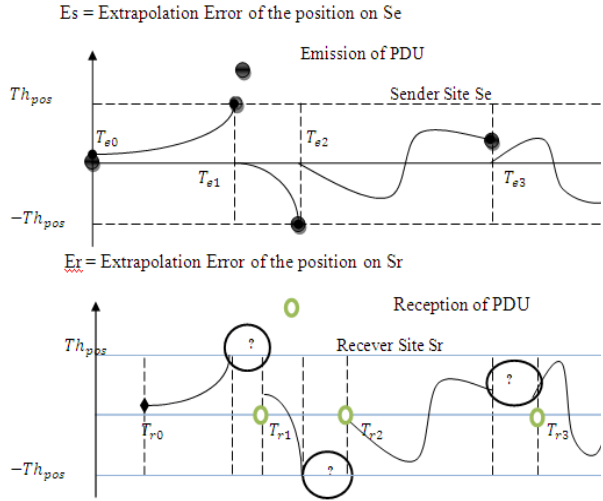

Figure 4: Evolution of the extrapolated error on both the sender site Se and the receiver site Sr.

We analyze the error at the both the sender site Se and the receiver site Sr:

<u>In the sender site $S_e$</u>:
- At $T_{e0}$, the extrapolation error of the position of the entity *A* increases and attains its maximum its maximum value (Thpos or DR) at Te1: at that time the high fidelity DR model begins sending data packets to correct the path with the right value of the position of A and the error Er become near to zero.
- From Te1 the previous scenario is reproduced to date Te2.
- At Te2, the error oscillates between Thpos and –Thpos without leaving this interval, during the HEART BEAT TIMER, that to say Te3 = Te2 + 5s. The DR model in the sender site Se transmits packets to bring error to zero.

<u>In the receiver site $S_r$</u>:
At $T_{r0}$, the extrapolation error Er remains the same as in site Se at the same time Te0; particularly, the error Er reach the maximum allowed value (Thpos at Te1, the date when refresh packets are send). However, the update of the position of A is made at time Tr1. The interval between [Te1, Tr1] corresponds to the network delay, where packets move from the sender to the receiver. In fact, it seems within this interval the error Er is not predictable and can exceed the Dead Reckoning threshold (Thpos) generating the spatial incoherence.

Generally, we highlight the lack of control of the error Er during the interval marked in interrogation point.

These analyses show that the actual definitions of the QoS guarantee of the spatial coherence during the simulation except in the interval [Te0, Tr0] separating the emission and the reception of update packets: during other periods the error Er exceed (the absolute value of $E_r$) transiently the maximum allowed value.

According to this analyze, two major questions can be asked: (1) does the spatial incoherence is prejudicial to the progress of a correct DIS exercise? (2) if so, how can we resolve this problem in order to control the transiently exceed of the error Er?

Concerning the first question, the transiently exceed of the error Er becomes detrimental when the latency DT is important ahead the period separating the reception of two consecutive refresh packets. If this is the case, the remote sites will have an incoherent spatial view during several times. This risk appears in large scale distributed networks.

To answer to the second question, we need to formulate a new investigation to fulfill the previous illustrated QoS requirements. Furthermore, in order to guarantee that Er does not exceed the maximum allowed value, we need to fix a maximum value of the error, what we call $E_{max}$ during the delay DTmax (see figure 5).

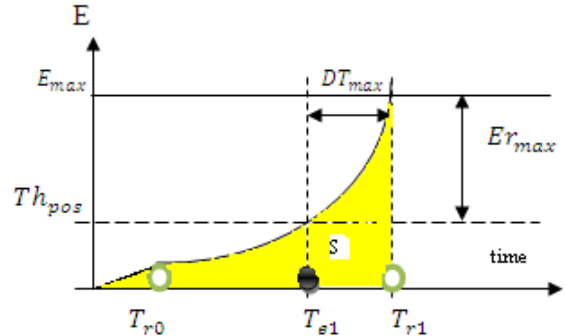

Figure 5: the transiently exceed of the error Er

In fact, the excess of the error is directly related to the dynamical behavior of the entity, as consequence the maximal allowed error $E_{rmax}$ is closely related to the maximum network delay $DT_{max}$ as shown in figure 5.

Let's analyze this situation in more closely way using some mathematical formulation.

## 5. Mathematical formulation of the error

In mathematics, a surface integral is defined integral taken over a surface. In our case, the surface S is in gray color like shown in figure 6.

Suppose we have a nice surface S and a function $f: S \to \mathfrak{R}$ defined on the surface. We want to define an integral of $f$ on S as the limit of some sort of Riemann sum.

To simplify the presentation, we assume the surface is sufficiently smooth to allow us to approximate the area of small piece of it by a small planar region, and then add up these approximations to get a Riemann sum. One way to calculate the surface integral is to split the surface to several small pieces $S1, S2, \ldots, Sn$ each having area $\delta S$, select points $r_i = (x_i, y_i) \in Si$, and form the Riemann sum:

$$R = \sum_{i=1}^{n} f(r_i) \cdot \delta S \qquad (2)$$

We take finer and finer subdivisions and the Riemann sums have a limit, and we call this limit the integral of $f$ on S (c.f relation ()):

Note that to find an explicit formula of this integral, we need to parameterize S by considering on S a system of curvilinear coordinates, and for example the parameterization be x(u, t), where (u, t) varies in some region D in the plane. Then, we need a vector description of S, say $r: D \to r(D) = S$. The surface is subdivided by subdividing the region $D \subset R^2$ into rectangles.

Let's look closely at one of the subdivisions:

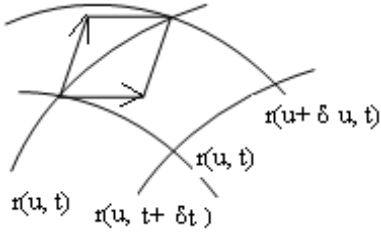

Figure 6: one subdivision area of S

We paste a parallelogram at the point $r(u_i, t_i)$ as shown in the above figure. The lengths of the sides of this parallelogram are:

$$\left| \left( \frac{\partial r}{\partial u}(u_i, t_i) \right) \partial u \right| \text{ and } \left| \left( \frac{\partial r}{\partial t}(u_i, t_i) \right) \partial t \right| \qquad (3)$$

The area is then is given by the following relation:

$$\left| \left( \frac{\partial r}{\partial u}(u_i, t_i) \right) \partial u \times \left( \frac{\partial r}{\partial t}(u_i, t_i) \right) \partial t \right| \qquad (4)$$

and we use the approximation

$$\partial S \approx \left| \left( \frac{\partial r}{\partial u}(u_i, t_i) \right) \times \left( \frac{\partial r}{\partial t}(u_i, t_i) \right) \right| \partial u \partial t \qquad (5)$$

and the Riemann sum:
$$R = \sum_{i=0}^{n} f(r(u_i, t_i)) \times \left| \left( \frac{\partial r}{\partial u}(u_i, t_i) \right) \left( \frac{\partial r}{\partial t}(u_i, t_i) \right) \right| \partial u \partial t \qquad (6)$$

These are the Riemann sums for the usual old time double integral function:
$$F(u, t) = \sum_{i=0}^{n} f(r(u_i, t_i)) \times \left| \left( \frac{\partial r}{\partial u}(u_i, t_i) \right) \left( \frac{\partial r}{\partial t}(u_i, t_i) \right) \right| \qquad (7)$$

So, within the plan D we can get the formula:

$$\iint_S f(r) dS$$
$$= \iint_D f(r(u_i t_i)) \left| \left( \frac{\partial r}{\partial u}(u_i, t_i) \right) \left( \frac{\partial r}{\partial t}(u_i, t_i) \right) \right| dS$$

$$(8)$$

Proof:
Lets a function f integrable on [u, δu], bounded and f is extendable by continuity in $\delta u^-$. This implies that f has a finite limit in $\delta u^-$. So, the integral $\int_u^{\delta u} f(t) dt$ converge to finite limit that we note L.

More generally, let's assume $(U, F, \mu)$ a measured space and E a metric space and a complex function f defined in $E \times U$. So, if for all $t \in E$ we have:
- The function $u \to f(u, t)$ is measurable.
- and f is continued on E,
- Then it exists a function $g \in L1$, so that for all $u \in X, |f(u, t)| \leq g(u)$.  (9)

Let's use our new-found knowledge and the extrapolation equation founded in the equation (1) to find the area of our Dead Reckoning scheme: note $P_a(t)$ the effective position curve and $P_{DR}(t)$ the extrapolated motion curve.

Note that the motion of the entity is independent from the choice of the surface parameters, only depends on the orientation of the velocity (changing the orientation implies the change of its sign).

We note the gap between the effective curve and the DR curve.

$$E_p(t) = \|P_a(t) - P_{DR}(t)\| \qquad (10)$$

We demonstrated, on one hand, that the surface S can be written with the surface integral formula (equation ()), and this function is approximated the Riemann sum, so it satisfied the Riemann condition. On that other hand, we can assume that this function is monotonic and convergent.

As reminder, within a Normed Vector Space (**E, ‖ ‖**), we can always define in canonical form a distance d from its norm. Just we pose:

$$\forall (x, y) \in E \times E, d(x, y) = \|y - x\| \quad (11)$$

We can also use the variational form of the Euclidean distance between two points A and B defined by $A = \vec{r}(u)$ and $B = \vec{r}(v)$ and we get the following:

$$D = \int_u^v dt \sqrt{\left( \frac{\partial \vec{r}(t)}{\partial t} \right)^2} \qquad (12)$$

The Euclidean distance between two points can be also generalized to the case of the objects and can be written like this:

$$D = \int_u \int_t dudt \left\{ \sqrt{\left(\frac{\partial \vec{r}(u,t)}{\partial t}\right)^2} + \lambda \left[\sqrt{\left(\frac{\partial \vec{r}(u,t)}{\partial t}\right)^2} - 1\right]\right\} \quad (13)$$

Where $\lambda$ is a Lagrange multiplier and its role is to ensure that the length remains the same during the transformation.

Also, using the triangle inequality in Normed Vector Space V we can get the relation (14):

$\|x + y\| \leq \|x\| + \|y\| \; \forall \, x,y \in V$ \quad (14)

From the above discussion, especially from the relations (8), (9),(12) and (14), we can observe the existence of real number $E_{max}$ which implies the following:

$$\begin{aligned}
E_p(t) &= \|P_a(t) - P_{DR}(t)\| \\
&= \left\|\int_{T_{ei}} du \int_{T_{ei}}^u [A_a(t) - A_i(t)]d\tau\right\| \\
&\leq \int_u du \int_{T_{ei}}^u |A_a(\tau) - A_i|d\tau \; + \\
&\quad \int_u \int_{T_{ei}}^{T_{ei+1}} [A_a(t) - A_i(t)]d\tau \; + \\
&\quad \int_{T_{ei+1}}^{T_{ei+1}+DT} \int_{T_{ei+1}}^u [A_a(t) - A_i(t)]d\tau \\
&\leq E_{max} \quad (15)
\end{aligned}$$

Note $A_a(t)$ and $A_i(t)$ are respectively the effective instantaneous acceleration of the simulated entity at time (t) and its acceleration at time $T_{ei}$.

Note that the error can be approximated to the distance D. So, we can calculate this distance using three terms in equations (11) and (13): (1) the DR Threshold (Thpos in our case, just to simplify) (2) the norm of the acceleration (considering any acceleration vector in the curve) and (3) the term $(V(t) - V(T_{ei+1}))DT$.

Thus, if we can fix the Threshold value and we can the norm of the acceleration vector, we will need just to obtain a value of the third term. What we need is to choose the best value of the Lagrange multiplier $\lambda$ which provides a way to resolve our optimization problem; technically, the Lagrange multiplier corresponds to the point where the differential of the function $f(u,t)$ has an orthogonal kernel to the gradient to the subject function $g(u,t)$ in this point.

So, in order to optimize the new-founded value of $Emax$, we will proceed to an artificial intelligent based mechanism to minimize the maximum allowed error which satisfies the spatial and the temporal coherence of the simulation. This method is referred to us Adaptive Neuro-Fuzzy Inference System (ANFIS) Reckoning. Results and proofs found in this section are injected in the next section to be used within the fuzzy inference system.

## 6. Adaptive Neuro-Fuzzy Inference System based Dead Reckoning

Each technique of the Artificial intelligence has particular properties that make them more adequate for a giving problem. For example, Neuronal Networks are interesting for the recognition, but are very bad for the decision making. The Fuzzy Logic which judge vague and imprecise are very adapted to decision making however it cannot automate the choice of the decision rules.

These shortcomings were the primary motivator to create a hybrid intelligent system where both the two technique are combined to overcome their individual limits.

Therefore, a Nero-Fuzzy Dead Reckoning algorithm based on the Adaptive Nero-Fuzzy Inference System (ANFIS) is studied to provide more flexible scheme to fulfill the required QoS in Distributed interactive simulation applications discussed early in this paper.

### 6.1. The ANFIS System

The proposed algorithm is referred to as ANFIS [12], which stands for Adaptive Neuro-based Fuzzy Inference System. This technique is based on fuzzy a system which is trained by the learning algorithm derived from the neuronal network theory. While the learning capacity is advantage from the viewpoint of Fuzzy Inference System (FIS), the formation of linguistic rule base will be advantage from the viewpoint of Artificial Neural Networks (FNN). Intelligent control of distributed entities is viable alternative to conventional model based DR control algorithm.

The introduced algorithm takes all proprieties of entities into consideration when measuring the behavior of the entity then it ensures the evaluation of this behavior is real time.

### 6.2. The ANFIS Dead Reckoning Model

For simplicity, the ANFIS DR Model considered in our case in composed of three inputs (a1, a2, a3) and only one output calculated from these inputs. The mapping output function f is given by:

$f^k = f(x^k) = f(a_1^k, a_2^k, ..., a_n^k); k \in 1..K$ \quad (16)

The learning process associates instantaneously the three inputs to the output in order to converge to the finally estimated output as shown in equation (17)

$\{(a1^1 a2^1 a3^1 f^1), \dots, (a1^k a2^k a3^k f^k)\}$ (17)

The position, the velocity and the orientation properties were employed in the simulation. The orientation includes the view direction in the interval $\left[-\frac{\pi}{2}, \frac{\pi}{2}\right]$, the position was chosen regarding the fixed threshold in the interval $[-Th_{pos}, Th_{pos}]$ and also the velocity interval is given in the interval $[-V_{max}, V_{max}]$.

From the last discussion in section 5, we found that we need to optimize the error regarding three terms: the threshold, the acceleration and the velocity and what we need is to estimate the new updated position with the respect of minimal admissible error.

In order to describe the linguistic terms of the Fuzzy premises, the membership functions and the linguistic terms we used the Sugeno [13] IF "…" THEN rules:
 ℝ1 = IF a1 is L1 and a2 is L2 and a3 is L3 then f = $z_{m1}$
 ℝ2 = IF a2 is H1 and a2 is H2 and a3 is L3 then f = $z_{m2}$
 ℝ2 = IF a1 is H1 and a2 is H2 and a3 is H3 then f = $z_{m3}$

Where i = 1...n and *Li and Hi* are the membership degrees of each fuzzy membership function and $Z_m$ are real values.

Every linguistic variable can have seven linguistic terms:{NB, NM, NS, ZE, PS, PM, PB} (*the training sets derived from this terms can be written in the form Negative Big, Negative Medium, Negative Small, Zero, Positive Small, positive Medium, Positive Big*) and their membership function are of the sigmoid form characterized by three parameters:

$sigmoid(x; a, b, c) = \frac{1}{1+\exp[-a(x-c)]}$ (18)

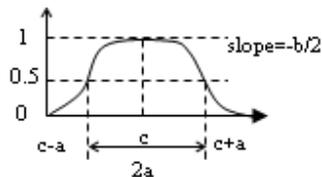

Figure 7: Meaning of parameters in generalized bell function

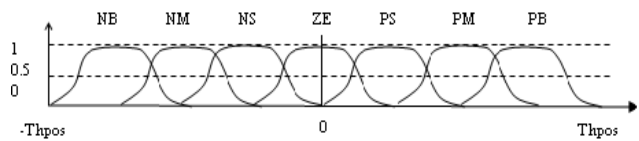

Figure 8: Initial linguistic terms for the input variables

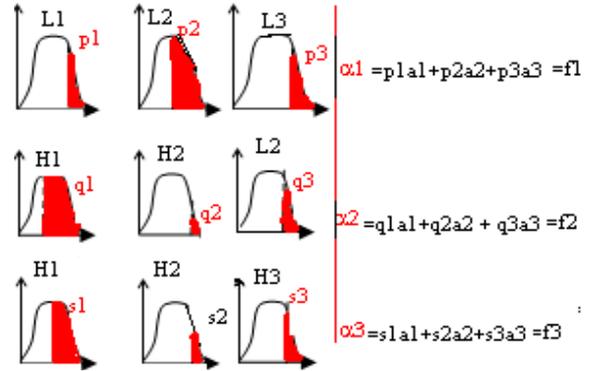

Figure 9: fuzzy Inference system of Takagi-Sugeno

For example, when the position of the entity is in close proximity to *Thpos* the value of the position is taken from the membership function of each curve within its definition interval, and so one. The fuzzy values are given in table 1. These values are not fixed by the user, but the ANFIS algorithm calculates them during the simulation and applies the value needed at each step of the simulation.

Table 1: Fuzzy members calculated at the fuzzy layer

| Distance (m) | Fuzzy members | | | | | | |
|---|---|---|---|---|---|---|---|
| | NB | NM | NS | ZE | PS | PM | PB |
| -Thpos | 85% | 65% | 0% | 0% | 0% | 0% | 0% |
| -2Thpos/3 | 25% | 70% | 40% | 0% | 0% | 0% | 0% |
| -Thpos/3 | 0% | 35% | 70% | 10% | 0% | 0% | 0% |
| 0 | 0% | 0% | 30% | 90% | 30% | 0% | 0% |
| Thpos/3 | 0% | 0% | 0% | 25% | 90% | 30% | 0% |
| 2Thpos/3 | 0% | 0% | 0% | 0% | 30% | 85% | 35% |
| Thpos | 0% | 0% | 0% | 0% | 20% | 40% | 90% |

Indeed, the ANFIS algorithm estimates continually the information (position, orientation, direction, velocity…) sent from each remote simulator and at each step it can got an idea about the behavior of the simulated object. So the decision making is taken very faster.

The ANFIS network used in this investigation was a five layer networks (figure), with six nodes in the first layer representing the each dimension of the input vector, one node in the last layer representing the output, and 3 hidden layers consisting of three nodes in each layer. This network attempts to develop a matching function between the input and the output vectors by using some training algorithms.

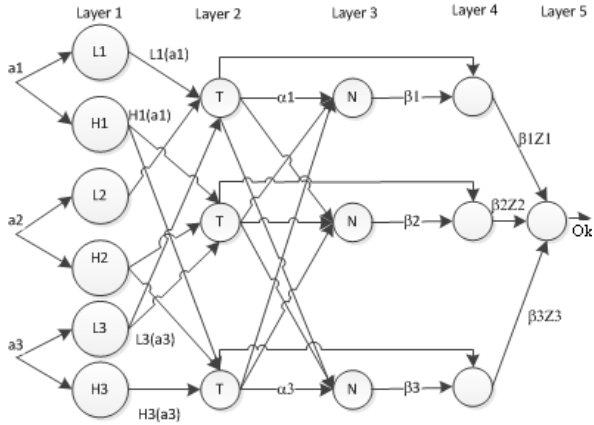

Figure 10: ANFIS Architecture of the DR Algorithm

Figure 10 describes the ANFIS architecture of the DR Algorithm. Each layer in this architecture has specific role:

**Layer 1:** the output of each node is the degree to which the given input satisfies the linguistic term associated to this node. The values of the fuzzy members (Table 1) are calculated in this layer.

**Layer 2:** each node realizes a T-norm function to compute the firing rules. The outputs of these nodes (called rule nodes) are given in relation (1)

$$O_{i,j} = \alpha_i = \mu_{Li}(a1).\mu_{Li}(a2).\mu_{Li}(a3) \quad (19)$$

**Layer 3:** the outputs of the T-norm are normalized in this layer. The output f the top, middle and bottom neuron is the normalized firing level of the corresponding rule:

$$\begin{cases} \beta_1 = \frac{\alpha_1}{\alpha_1+\alpha_2+\alpha_3} \\ \beta_2 = \frac{\alpha_2}{\alpha_1+\alpha_2+\alpha_3} \\ \beta_3 = \frac{\alpha_3}{\alpha_1+\alpha_2+\alpha_3} \end{cases} \quad (20)$$

**Layer 4:** the learning process is done in this layer. The instantaneously calculated output (in the case of on line learning process) is compared to the estimated output. If the off line learning technique is used, the estimated output is stored in database and then compared the output of the algorithm. Le output of each neuron is given by the normalized product of the firing rule and the output correspondent rule, like shown in relation (20):

$$\begin{cases} \beta_1 z_1 = \beta_3 \times \alpha_1 \\ \beta_2 z_2 = \beta_3 \times \alpha_2 \\ \beta_3 z_3 = \beta_3 \times \alpha_3 \end{cases} \quad (21)$$

**Layer 5:** this is a single node of the output function of the overall system, as the sum of all incoming signals, i.e.

$$O_k = \beta_1 z_1 + \beta_2 z_2 + \beta_3 z_3 \quad (22)$$

Rather than choosing the parameters associated with a given membership function arbitrarily, these parameters could be chosen so as tailor the membership functions to the input/output data in order to account for these types of variations in the data values. The Neuro-adaptive learning method works similarly to that of neural networks.

The ANFIS method provides a method for fuzzy modeling procedure to learn information about a data set; it constructs a fuzzy inference system (FIS) whose membership function parameter are tuned (adjusted) using either the back-propagation algorithm (gradient descent method in equation (22)) alone or in combination with the last square error type of method (equation (23)), then it maps inputs through input membership functions and associated parameters, and through the output membership functions and associated parameters to outputs it can interpret the input/output map.

$$b_i(t+1) = b_i(t) - \eta \times \frac{\partial E}{\partial b_i} \quad (23)$$

Where $\eta$ is the learning rate of the gradient method.
The parameters associated with the membership functions changes through the learning process. The computation of these parameters (adjustment) is facilitated by the gradient vector. This gradient vector provides a measure of how well the fuzzy inference system is modeling the input/output data for a given set of parameters. When the gradient is obtained, the optimization routine is applied in order to adjust the parameters to reduce some error measure. This is defined by the squared combination of the least squared error between the actual and the desired output given in the equation (22).

$$E_k = \frac{1}{2}(y_k - o_k) \quad (24)$$

Where $y_k$ is the $k^{th}$ component of the $p^{th}$ desired vector and $O_k$ is the $k^{th}$ component of the actual output vector produced by presenting the $p^{th}$ (Layer) input vector to the network.
From the relation (6), (7) and (8) we founded a way to calculate the surface as Riemann sum, as consequence we can easily found the error measure E for the $p^{th}$ of the training data as the sum of the squared error in all input layers:

$$E = \sum_{p=1}^{P} E_k \qquad (25)$$

The question is when we can choose the best value of the admissible error. Obviously, when the error $Ep$ is equal to zero, the network can is able to bear all the traffic required for the consistency of the simulation and provides the required throughput. In that case, the ANFIS model succeeds in reducing both the spatial coherence and the temporal coherence associated with remote entity.

Since we subdivided the region S into several infinitesimal parties, each one representing the local error calculated from the output of the ANFIS model, we used each subdivision to distribute the membership functions of the inputs of the Fuzzy Inference System (FIS). Thus, the selection of the membership value for each input is done by the gradient descent method. Therefore, each membership value is well located within the curve (see figure 6).

In order examine the power of the ANFIS DR model; we developed a C++ library which provides the computation tools used in the simulation.

Table 2: Extrapolation Error for three different approaches

| Prediction Horizon | ANFIS | Neuro-Reckoning |
|---|---|---|
| 1 | 0.317 | 0.356 |
| 2 | 0.313 | 1.468 |
| 3 | 0.312 | 2.02 |
| 4 | 0.310 | 2.2936 |
| 5 | 0.309 | 3.67 |
| 6 | 0.306 | 5.138 |
| 7 | 0.305 | 6.606 |
| 8 | 0.304 | 8.074 |
| 9 | 0.301 | 9.542 |
| 10 | 0.30 | 11.01 |

The results were compared to other some other approaches taken from the literature (see Table 2). The Neuro-Reckoning was used as predictive contract mechanism was presented is [15]. From the inspection of the table, we can note that ANFIS DR model is slightly better results to predict the best value of the threshold error. Furthermore, the distance between the objects (see equation (13)) remains minimal when the Lagrange multiplier is very close to the gradient learning rate updated by the gradient descent method given in equation (23).

Of a particular interest is the decrease of the value of the extrapolated error in the receiver site.

## 7. Conclusion

This paper suggests an overview of the Dead Reckoning Algorithms used in distributed interactive communication and focus on the QoS fulfillment in large scale networks. Then we proposed a Neuro-Fuzzy Dead Reckoning protocol which provides more flexible scheme to fulfill the required QoS. It uses the benefits of Fuzzy Inference and the Neural Networks: Fuzzy logic can encode the expert knowledge directly using rules with linguistic labels, and then these quantitative labels are injected within the learning process of Neural Networks which automated this process using the back-propagation algorithm and improved the performance of the developed scheme. The overload of communication is lean and the bandwidth reducing is enhanced. The extrapolated error was minimized according to the mathematical approach presented here. The simulation results validate the potential of our proposed solution and explicit the efficiency of our decomposition scheme to improve the predictive performance.

Many more ideas, protocols and products have implemented some kind of DR algorithms. However, it is still a matter of research to find out which involve the QoS management over predictive protocols in distributed asynchronous communication networks.

## 8. Acknowledgment

The proposed work is involved within the PLATSIM Project supported both by the LAAS-CNRS and ECA FAROS. Authors would like to think the partner of this project.

## Author Biographies

**AKRAM HAKIRI** is a Ph. D student in the University of Toulouse and researcher in the LAAS-CNRS French research Labs, Toulouse-France. He has his master degree from the University of Paul Sabatier in Toulouse, France and he worked in Wireless Sensor Networks for spatial and Aeronautic systems. He is also an engineer in computer science and automatics from the National Institute of Applied science and Technology (INSAT) in Tunisia.

**PASCAL BERTHOU** is an associate professor in computer science in the University of Toulouse and researcher in the LAAS-CNRS French research Labs, Toulouse-France. He worked in network support for the distributed interactive simulation, wireless sensor networks, multi-network communication architecture and multimedia applications over broadband satellite systems.

**THIERRY GAYRAUD is** Full Professor in the University of Science, Toulouse III, France and researcher in the LAAS-CNRS French research Labs, Toulouse-France. His research interests are sensor networks, QoS in satellite communication system and QoS for distributed interactive simulation application.